\documentclass[prl,twocolumn, preprintnumbers,amsmath,amssymb]{revtex4}

\usepackage{amsmath,amssymb,eucal,graphicx,subfigure,bbding,booktabs,array,color}
\usepackage{mathrsfs}

\usepackage{graphicx}
\usepackage{dcolumn}
\usepackage{bm}

\begin{document}

\title{Deterministic endless collective evolvement in active nematics}

\author{Xia-qing Shi}\affiliation{National Laboratory of Solid State Microstructures,
Nanjing University, Nanjing 210093, China}\affiliation{Center for
Soft Condensed Matter Physics and Interdisciplinary Research,
Soochow University, Suzhou 215006, China}
\author{Yu-qiang Ma}\altaffiliation[ ]{Corresponding author. E-mail:
myqiang@nju.edu.cn.} \affiliation{National Laboratory of Solid
State Microstructures, Nanjing University, Nanjing 210093, China}
\affiliation{Center for Soft Condensed Matter Physics and
Interdisciplinary Research, Soochow University, Suzhou 215006,
China}

\begin{abstract}
We propose a simple deterministic dynamic equation and reveal the
mechanism  of large-scale endless evolvement of spatial density
inhomogeneity in active nematic. We determine the phase regions
analytically. The interplay of density, magnitude of nematic
order, and nematic director is crucial for the long-wave-length
instability and the emergence of seemingly fluctuated collective
motions. Ordered nematic domains can absorb particles, grow and
divide endlessly. The present finding extends our understanding of
the large-scale and seemingly fluctuated organization in active
fluids.
\end{abstract}

\maketitle

Assemblies of active particles, which may be a physical
abstraction of running animals\cite{bird1}, flying
birds\cite{bird2}, swimming bacteria\cite{bact1}, migrating
cells\cite{cell1} or even cytoskeleton\cite{cytoskeleton2}, have
been served as a new building block for physicists over the last
decade or so to understand the common collective behavior in these
non-equilibrium systems\cite{vicsek1}.  Active nematic, a recently
proposed concept for a kind of apolar active particles, is formed
by driven rod-like particles with head-tail symmetry  through the
randomly driving force along the rod orientation axis at the
single-particle level[7(a)-7(f)].  The symmetry of the system is
not broken by applying such micro-driven forces until spontaneous
symmetry breaking occurs. Recently, simulations and experiments in
active nematic system show well-organized collective motions  with
system-sized fluctuation[7(b)-7(e)]. For example, splitting and
merging of large-scale self-organized structures are exemplified
in the simulation on active nematics[7(c)]. Experiments also show
large-scale collective swarming and swirling in driven granular
rods monolayer[7(d),7(e)]. These observations lead us to think
about the nature of such seemingly fluctuated collective motions.
It is currently unclear whether these large-scale collective
motions arise in a deterministic manner or as a result of noises
applied upon the system. Moreover, are they genuinely restless on
large scale and evolving without end? These important aspects are
still not well addressed in previous studies.

\vspace{-6pt}

In the present study, we start from a deterministic equation to
study the mechanism of restless collective evolution in active nematics. We reveal a new phase that is characterized by the unattainability of stable steady state. We
first identify that, if the steady state can be reached, the system
investigated here favors spatially homogeneous state. On the other
hand, by taking account of the interplay between particle density
and local nematic order (i.e., magnitude and orientation), the
linear stability analysis shows that homogeneous nematic state can
be unstable to fluctuations of small wave number. Therefore, the
system enters into a chaotic phase region with no stable steady state. Large-scale spatial
inhomogeneity of density and nematic order  is developed as a
result of long-wavelength instability. The spatial inhomogeneity
in turn changes the direction of the nematic director, leading to
a non-ending evolvement of the system. Numerical flux analysis
shows that the particle-rich nematic domains are surrounded by
particles fluxes, and evolve via absorbing particles from
low-density isotropic medium, growing, and extending itself and
breaking into small pieces.
More importantly, all these seemingly fluctuated
collective motions giving rise to giant number fluctuations are
governed by a deterministic equation   which is  essentially free
of noises.

We notice that one salient
feature of simulation rules for active nematics by Chate et
al.[7(c)] is that particle rotations are governed through
inter-particle nematic interaction while spatial translational
movements are free of such interactions. Experimentally particles
are driven along their long axis, inducing strong longitudinal
diffusion, and they can thrust into the media with the supply of
kinetic energy[7(e)].  A simple diffusion equation which
follows these observations can be written as(see \cite{supp}):
 \begin{equation}
\begin{aligned}\label{dyadd11}
&\partial_t f(\textbf{r},\textbf{u},t) =\nabla[D_{\|}{\bf u}{\bf
u}\nabla+D_{\perp}(\textbf{I}-{\bf u}{\bf u})\nabla] f({\bf
r},{\bf
u})\\
&+\mathscr{R}[D_r\mathscr{R}f({\bf r},{\bf
u})+D_r\mathscr{R}w({\bf r},{\bf u})f({\bf r},{\bf u})],
\end{aligned}
\end{equation}
  where $D_{\|}$ and $D_{\perp}$ are the parallel and
perpendicular components of the translational diffusion constants.
$D_r$ is the rotational diffusion constant, and the rotational
operator $\mathscr{R}$ is defined by $\mathscr{R} ={\bf
u}\times\partial_{\bf u}$\cite{Shimada}.
$f(\textbf{r},\textbf{u},t)$ is the particle number distribution
function where the spatial coordinate ${\bf r}$   and the  unit
vector ${\bf u}$  denote the center-of-mass position and long-axis
direction of particles, respectively. $w(\textbf{r},\textbf{u})$
is a self-consistent interacting potential which has $\pm {\bf
u}$-symmetry. In two-dimensional(2-D) case, the most common form of such interacting potential
is the excluded-volume-like interaction
$w(\textbf{r},\textbf{u})=l^2\int{\rm d}{\bf u}'|{\bf u}\times{\bf
u}'|f({\bf r},{\bf u}')$, where $l$ is the particle length.

\begin{figure}
\includegraphics[width=7cm,bb=40 15 281 213]{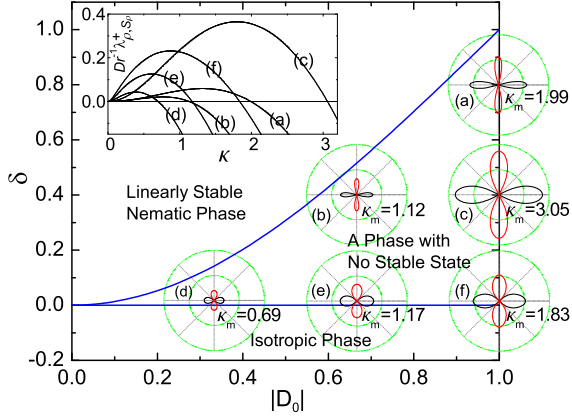}
\caption{Three phases are separated by the blue curves. Polar plots (a)-(f) indicate the instability regime for $\kappa$ and $\theta$.
 The locations ($D_0$, $\delta$) of their origins are used as the parameters
to produce these plots.  (a) $|D_0|=0$, $\delta=0.8$, (b) $|D_0|=2/3$, $\delta=0.4$, (c)
$|D_0|=0$, $\delta=0.4$, (d) $|D_0|=1/3$, $\delta=0.01$, (e) $|D_0|=2/3$, $\delta=0.01$,
and (f) $|D_0|=0$, $\delta=0.01$. The black and red branches in polar plots represent the
cases of positive and minus $D_0$, respectively. The inset
shows the corresponding instability modes
$D_r^{-1}\lambda^+_{\tilde{\rho},S_{\rho}}|_{\theta=0}$ for these polar plots.}
\end{figure}

The diffusion equation Eq.\eqref{dyadd11} for active nematics
satisfies particle number conservation with the spatial
translational current   ${\bf J}^s({\bf r},{\bf u})=-D_{\|}{\bf
u}{\bf u}\nabla f({\bf r},{\bf u})-D_{\perp}(\textbf{I}-{\bf
u}{\bf u})\nabla f({\bf r},{\bf u})$ and the local rotational
current ${\rm J}^r({\bf r},{\bf u})=-D_r\mathscr{R}f({\bf r},{\bf
u})-D_r\mathscr{R}w({\bf r},{\bf u})f({\bf r},{\bf u})$, which are
independent of each other. The translational current is purely
diffusive. In the Fourier space as defined by $f({\bf r},{\bf
u})=\int{\rm d}{\bf r}\tilde{f}({\bf k},{\bf u}) e^{-i{\bf
k}\cdot{\bf r}}$, the spatial fluctuation modes are governed by
diffusive decaying term $-k^2(D_{\|}{\rm
cos}^2\varphi+D_{\perp}{\rm sin}^2\varphi)\tilde{f}({\bf k},{\bf
u})$, where $\varphi$ is the angle ${\bf k}$ makes with ${\bf u}$.
For positive $D_{\|}$ and $D_{\perp}$, all the spatial fluctuation
modes decay except for $k=0$ which indicates total particle number
conservation. Therefore, as governed by these decaying diffusive
modes, the spatial term suggests that only the spatially
homogeneous state will probably be the stable steady state if
there are no particle sources and sinks in the system and at the
boundaries. However, when $D_{\|}\neq D_{\perp}$, we will show that the spatially homogeneous state may become unstable to fluctuations in the nematic state. Consequently, the system is deprived of all possible stable steady state and
becomes restless and evolves endlessly, similar
to the deterministic nonperiodic flow found in
turbulence\cite{Lorenz}.

We first examine the spatially homogeneous dynamic
equation derived from Eq.\eqref{dyadd11} by neglecting the spatial derivatives: $
\partial_t
f(\textbf{u},t)=\mathscr{R}[D_r\mathscr{R}f({\bf u})+D_r\mathscr{R}w({\bf u})f({\bf
u})].$ Determined by the integration kernal of the self-consistent
interacting potential $w(\textbf{u})$, $f(\textbf{u},t)$ has $\pm
{\bf u}$-symmetry. In spatially homogeneous nematic state, we
assume that the nematic director ${\bf n}_0$ is in the $x$-axis, and
thus the distribution function can be expanded as
$f(\textbf{u},t)=(2\pi)^{-1}\rho\sum_{n=0}^\infty a_n(t) {\rm
cos}2n\phi$, where $\rho$ is the particle number density, $\phi$
is the angle of the unit vector ${\bf u}$, and $n=0,1,2\cdots
\infty$. The dynamic equation of the coefficient $a_n(t)$ is given by
${(4D_r)}^{-1}{\partial_t
a_n(t)}=-n^2a_n+\frac{\rho l^2 n^2a_n}{(4n^2-1)\pi}+\rho
l^2\sum_{m=1}^\infty\frac{nma_m(a_{|n-m|}-a_{|n+m|})}{(4m^2-1)\pi}$,
 where $a_0(t)=1$ and $a_1(t)=2 S(t)$, since the number density  $\rho=\int {\rm
d}{\bf u} f({\bf u},t)$ and the nematic order parameter $S(t)=\int
{\rm d}{\bf u}{\rm cos}2\phi f({\bf u},t)/\rho$.
 By setting $a_n=0$ for $n\geq 3$,  the truncated dynamic equation for the
nematic order parameter can be written as:
$\frac{\partial_tS(t)}{4D_r}=(\tilde{\rho}-1)S-\frac{3\tilde{\rho}^2S^3}{4(5-\tilde{\rho})}$,
where $\tilde{\rho}=\rho/\rho^* $ is the rescaled number density, and the critical
density $\rho^*=3\pi/2l^2$, beyond which the system enters into a spatially homogenous
nematic state.


Next, we examine the linear stability of such spatially
homogeneous nematic state above $\rho^*$, by expanding  the number
distribution function $f({\bf r},{\bf u})={(2\pi)^{-1}}\rho({\bf
r})[1+4({u}_{\alpha}{u}_{\beta}-\delta_{\alpha\beta}/2)Q_{\alpha\beta}({\bf
r})]$
 with the
inclusion of the alignment tensor  $Q_{\alpha\beta}({\bf
r})=S({\bf r})(\hat{n}_\alpha\hat{n}_\beta({\bf
r})-\delta_{\alpha\beta}/2)$ where $\hat{\bf n}(\bf r)$ is the
unit vector of the nematic director. We assume that nematic
director is along the $x$ axis of the system. Small fluctuations
of density and nematic director near the ordered nematic state are
given by $\delta\tilde{\rho}({\bf r})=\tilde{\rho}({\bf
r})-\tilde{\rho}_0$ and $S\delta n_y({\bf r})=Q_{xy}({\bf r})$,
respectively. Here, $\tilde{\rho}_0=\rho_0/\rho^*$ is the reduced
bulk particle density, and $\delta n_y({\bf r})$ is the
$y$-component of the deviated nematic director.  Noting that ${\bf
n}_0({\bf r})=\hat{{\bf x}}$ and $|{\bf n}|=1$,  $\delta n_y({\bf
r})$ is the only possible small fluctuation of nematic director
${\bf n}({\bf r})$. The resulting hydrodynamic equations can be
obtained from Eq. \eqref{dyadd11}, yielding
\begin{eqnarray}
\label{density1}\partial_t\delta\tilde{\rho}\!\!&=&\!\!\frac{D_p}{2}\partial_\alpha^2\delta\tilde{\rho}+\frac{D_n}{2}(\partial_x^2-\partial_y^2)(\tilde{\rho}S)\nonumber\\
&&+2D_n\tilde{\rho}_0S\partial_x\partial_y\delta n_y,\\
\label{nematicd1}\tilde{\rho}_0 S\partial_t\delta
n_y\!\!&=&\!\!\frac{D_n}{4}\partial_x\partial_y\delta\rho+\frac{D_p}{2}\tilde{\rho_0}S\partial_\alpha^2\delta n_y,\\
\label{orderp1}\partial_t[\tilde{\rho}S({\bf
r})]\!\!&=&\!\!\frac{D_n}{4}(\partial_x^2-\partial_y^2)\tilde{\rho}+\frac{D_p}{2}\partial_\mu^2(\tilde{\rho}S)\nonumber\\
&&+4D_r\tilde{\rho}(\tilde{\rho}-1)S-D_r\frac{3\tilde{\rho}^3S^3}{5-\tilde{\rho}},
\end{eqnarray}
where $D_p=D_{\|}+D_{\perp}$ and $D_n=D_{\|}-D_{\perp}$.
It is easy to see that the homogeneous state is stable to fluctuations of coupled nematic director
and density field.
Here we consider the stability of the modes that couple fluctuations of density $\delta\tilde{\rho}({\bf r})=\tilde{\rho}({\bf
r})-\tilde{\rho}_0$ and magnitude of nematic order $\delta S_\rho({\bf r})=\tilde{\rho}({\bf
r})S({\bf r})-\tilde{\rho}_0S_0$ with $\delta n_y=0$ around the homogeneous state
$(\tilde\rho_0,S_0)$, where
$S_0=\sqrt{4(5-\tilde{\rho}_0)(\tilde{\rho}_0-1)/3\tilde\rho_0^2}$.
The mode of fluctuations in Fourier components with wave vector
${\bf k}$, defined by $\delta \tilde{\rho}({\bf r})=\int{\rm
d}{\bf k}\tilde{\rho}_{\bf k} e^{-i{\bf k}\cdot{\bf r}}$ and
$\delta S_{\rho}({\bf r})=\int{\rm d}{\bf k}S_{\rho{\bf k}}
e^{-i{\bf k}\cdot{\bf r}}$, is governed by
\begin{gather}\label{kspace}
\partial_t\!\!
\begin{bmatrix}
\tilde{\rho}_{{\bf k}}\\\\S_{\rho\bf k}
\end{bmatrix}
\!\!=\!-\frac{1}{2}\!\!
\begin{bmatrix}
D_pk^2 & D_n\cos\negmedspace 2\theta
k^2 \\\\
\begin{array}{ll} D_n\cos\negmedspace 2\theta k^2/2\\-8D_r\tilde{\rho}_0S_0
\end{array} & \begin{array}{ll} D_pk^2 \\+16D_r\delta \end{array}
\end{bmatrix}\!\!\!
\begin{bmatrix}
\tilde{\rho}_{{\bf k}}\\\\S_{\rho \bf k}
\end{bmatrix}
,
\end{gather}
where $\delta=\tilde{\rho}-1$, $\theta$ is  the angle between wave
vector ${\bf k}$ and nematic director ${\bf n}_0$. The eigenvalues
of the coefficient matrix in Eq.\eqref{kspace} are given by
$D_r^{-1}\lambda^{\pm}_{\tilde{\rho},S_{\rho}}=-(8\delta+\kappa^2)/2\pm
\sqrt{D_0^2\kappa^4\cos^2\negmedspace 2\theta/8-4\sigma D_0
\kappa^2\cos\negmedspace 2\theta+16\delta^2}$, where
$\sigma=\sqrt{(4-\delta)\delta/3}$, the rescaled   coefficient
$D_0=D_n/D_p$, and the wave number $\kappa=\sqrt{D_p/D_r}k$.
The real part of $\lambda^-_{\tilde{\rho},S_{\rho}}$ is always negative, representing
stable decaying mode. However, the mode $\lambda^+_{\tilde{\rho},S_{\rho}}$ becomes
positive when $
32(D_0\sigma\cos\! 2\theta+\delta)\kappa^2+(2-D_0^2\cos^2 \!
2\theta)\kappa^4<0$. The coefficient of $\kappa^4$ is always
positive since $|D_0|\leq 1$, signifying that for large enough wave numbers, the
fluctuations are always stable. For small wave numbers which describe large-scale
fluctuations, the stability is controlled by the coefficient of $\kappa^2$. Thus when
$(D_0\sigma\cos\negmedspace 2\theta+\delta)<0$, the system becomes unstable on large
scale.

 The phase map for $(D_0,\delta)$ is given in Fig. 1. Between the
isotropic and linearly stable nematic phases, there is a region
where spatially homogeneous nematic state is unstable.  It is
denoted as a `phase with no stable state' to emphasize that the
only possible form of steady-state solution is unachievable there.
For different $(D_0,\delta)$ within the `no stable state' region,
the instability mode structures $(\kappa,\theta)$ are given by the
polar plots whose central positions represent $(D_0,\delta)$.
Here, each polar plot is composed of horizontal (black) and
vertical (red) branches enclosing unstable fluctuation modes,
corresponding to $D_0<0$ and $D_0>0$, revealing that spatial inhomogeneities are developed parallel and perpendicular to the nematic director, respectively. The maximum
values $\kappa_m$ of  $\kappa$ for the instability regimes are
always in the directions $\theta=\pi/2,3\pi/2$ for $D_0>0$ and
$\theta=0,\pi$ for $D_0<0$, respectively. Generally speaking,
$\kappa_m$ becomes larger when $(D_0,\delta)$ is far from the
phase boundary.   The inset of Fig. 1 shows the value of
$D_r^{-1}\lambda^+_{\tilde{\rho},S_{\rho}}$, where for small
$\kappa$, $D_r^{-1}\lambda^+_{\tilde{\rho},S_{\rho}}
> 0$  corresponds to the long-wavelength instability and the onset of
large-scale spatial inhomogeneity.

\begin{figure}
\includegraphics[width=7cm,bb=2 0 48 42]{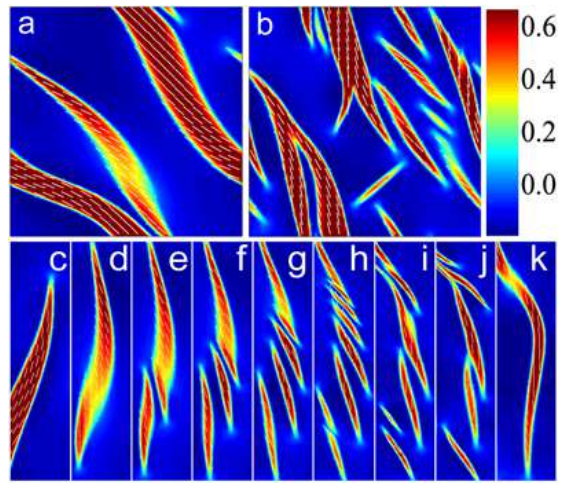}
\caption{Density and nematic order profiles are plotted. The color scale shows the local relative rescaled density value $\delta({\bf r})=\tilde{\rho}({\bf r})-1$. The length
and angle of white segments show the magnitude and direction of nematic order,
respectively. The dynamic parameters are $D_r=2$, $D_{\perp}=0.4$, and $D_{\|}=2.4$. The
reduced instability dynamic parameter $D_0=5/7$ and $\delta=0.01$. The system size is
$300 \times 300$  with the particle length $l=1$.  Periodic boundary condition is
implemented. Discrete time step $\Delta_t=0.018$ and spatial steps $\Delta_x=\Delta_y=\Delta_r=3$. (a)-(b) The snapshots are
taken at times $1.7\times 10^6\Delta_t$ and $2.3\times 10^6\Delta_t$.
(c)-(k) A close look of the breaking and coalescing processes of a nematic domain, at
times $1.6\times 10^6$, $1.7\times 10^6$, $1.71\times
10^6$, $1.72\times 10^6$, $1.73\times 10^6$,
$1.74\times 10^6$,  $1.75\times 10^6$,  $1.76\times 10^6$
and  $1.8\times 10^6$ in unit $\Delta_t$.}
\end{figure}


What will
happen in the phase region where there is no stable steady state? And how
the system evolves in time? To answer these questions we directly
integrate Eq. \eqref{dyadd11} numerically in this region using
alternative implicit algorithm(see \cite{supp}). Starting from an isotropic and
spatial-homogeneous initial condition, local ordered nematic
domains form at the beginning, accompanied with quick development
of density inhomogeneity. Further coarsening of these structures
leads to the coexistence of particle-enriched nematic domains and
particle-poor isotropic region where $\rho < \rho^*$ (Fig. 2a).
 However, such a large-scale spatially inhomogeneous structure
is unstable, and it will evolve and become fragmented as shown in
Fig. 2b. The fragmented structure will again coalesce and similar
process will repeat aperiodically and endlessly (see \cite{supp}
M1.mov). In Fig. 2c-2k, we show how a particle-rich nematic
branch breaks into pieces and re-unites into a structure with new
morphology.

\begin{figure}
\includegraphics[width=8.cm,bb=3 0 96 92]{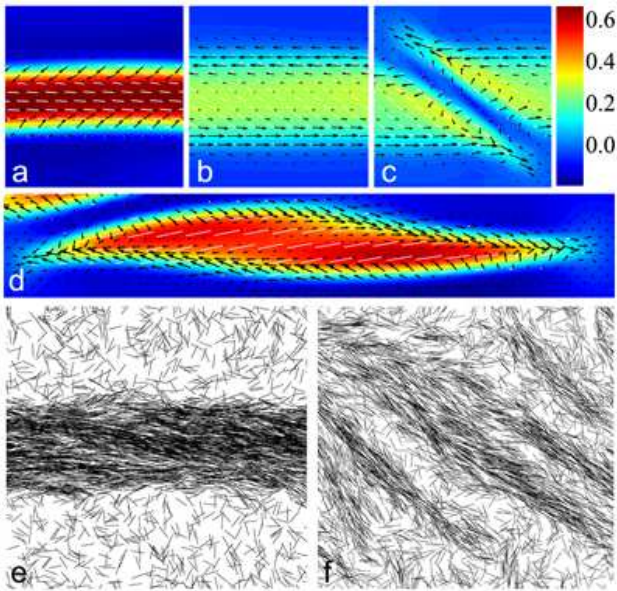}
\caption{(a)-(c) Three typical steps that an ordered stripe
breaks, at times $30000\Delta_t$,  $60000\Delta_t$ and
$63000\Delta_t$ in sequent. (d) Twisted spindle shaped structure
formed after it breaks off from a larger ordered structure.
 The color scale shows the local relative rescaled density
value $\delta({\bf r})=\tilde{\rho}({\bf r})-1$. The length and
angle of white segments show the magnitude and direction of
nematic order, respectively. The black arrow shows the direction
and strength of particle fluxes. (e)-(f) Simulation result shows
density inhomogeneity and similar restless evolvement in active
nematics for $9.4\times 10^4$ and $1.1\times 10^5$ simulation
sweeps.}
\end{figure}

How does the fragmentation process occur? In Fig. 3a-c, we take a
close look at the process that a nematic band breaks up(for a more
continuous process, see \cite{supp} M2.mov). In Fig. 3a, after the
spontaneous formation of a nematic band, initially, it is shown
that the nematic director in the high-density ordered region is
almost parallel to the density stripe boundary. In this case, the
nematic director is along the $x$-axis. For $D_n>0$, from our
previous stability analysis, the term
$-\frac{1}{2}D_n\partial_y^2(\rho S)$ in Eq.\eqref{density1} is
directly responsible for the development of such density
inhomogeneity. Now, we are interested if such aligned director
field is stable to small fluctuations $\delta {\bf n}_{\perp}({\bf
r})={\bf n}({\bf r})-{\bf n}_0=(0,\delta \hat{n}_{\perp y})$. To
linear order, the dynamic equation for $\delta \hat{n}_{\perp
y}({\bf r})$ can be obtained from Eq.\eqref{dyadd11} as
$\tilde{\rho}S\partial_t\delta n_{\perp
y}=\frac{1}{2}D_p[\tilde{\rho}S\partial_x^2\delta n_{\perp
y}+\delta n_{\perp y}\partial_y^2\tilde{\rho}S]$, where we have
assumed that there is no spatial variation of $\tilde{\rho} S$
along $x$-axis. The spatial variation
 of $\tilde\rho S$ along the $y$-axis is significant since density inhomogeneity is developed in that direction.
  Near stripe boundaries, we always have $\partial_y^2\tilde{\rho}S >0$,
 which makes the fluctuations $\delta n_{\perp y}$ unstable. Such instability will induce the
  change of the nematic orientation, and this explains
  why the nematic directors in Fig. 2 and Fig. 3b are most likely to be oblique to the
 density profile boundaries.  When the nematic directors become oblique to the boundary, as shown in Fig. 3b, there is a leakage of
 particles from the high-density region. As the particle density in the stripe falls into
 the `no stable state' region as shown in Fig. 1,
 the spatial instability takes place again. This leads to a fragmentation event, as shown in Fig. 3c. It is shown that a density crevice forms parallel with the nematic
 director as spatial instability requires. In Fig.~3d, we show a twisted-spindle
shaped high-density region with local nematic order, which is commonly formed after it breaks off from a larger ordered structure in Fig. 2e.

We also perform simulations to examine the stability of
homogeneous nematic state on the basis of  Eq.\eqref{dyadd11} (see \cite{supp}). Fig. 3e shows the formation of a
particle-enriched nematic band. We find that such band is also
unstable and undergoes similar process (see \cite{supp} M3.mov ),
where the nematic director changes its direction  in time and the
fragmentation event takes place afterward (Fig. 3f).


It is worth to notice that all
these highly dynamic structures are surrounded by particle fluxes
around the density profile boundaries. The density currents are
defined as ${\bf J}({\bf r})=(J_x({\bf r}),J_y({\bf r}))$, with
$J_\alpha({\bf r})=-\frac{1}{2}D_p\partial_\alpha\rho({\bf r})-D_n
\partial_\beta [\rho({\bf r}) Q_{\alpha\beta}(\bf r)]$  where the first term is just ordinary
diffusive current  and the second term is the current generated by
 coupling nematic directors.  As we show in Fig. 3a, inward
currents, generated by $(0,\frac{1}{2}D_n\partial_y(\rho S))$
which is included in the second term of ${\bf J}({\bf r})$, are
directly responsible for the development of density inhomogeneity.
As the time evolves,  in Fig. 3b, along the two sides of stripe
boundaries the system generates anti-parallel currents which also
originate from $-D_n\partial_\beta [\rho({\bf r})
Q_{\alpha\beta}(\bf r)]$, and the particles seem to move under a
self-organized rachet potential\cite{Julicher}.
 When the crevice forms in Fig.~3c,
particles flow into the low density region as guided by the nematic directors, accompanied with the growth of nematicly ordered tips
near the crevice. In Fig.~3d, as indicated by the density currents, the twisted-spindle
shaped nematic region absorbs
particles from the medium on both sides and generate outward flux on both
tips. In this way it can grow and extend itself quickly into the low-density medium. With the presence of these density fluxes around the ordered structure, it behaves like a creature absorbing, growing, dividing and dissipating into
isotropic medium endlessly.

 In summary, the dynamic equation abstracted from previous
simulations and experiments suggests a nematicly ordered phase
with no stable steady state. Thus the system must evolve
endlessly. We reveal the statistical mechanism that governs the
large-scale and seemingly fluctuated collective evolution. We show
that, as a result of long-wavelength instability, density and
order inhomogeneity develops as guided by nematic director field.
The spatial inhomogeneity in turn changes the directions of local
nematic directors. The changed nematic directors further guide the
fragmentation events, leading to endless evolution of the system.
Finally, it would be interesting to extend our analysis to other
active fluids.

This work was supported by the National Natural Science Foundation
of China (No. 10974080).


\begin{thebibliography}{10}

\bibitem{bird1} J.K. Parrish and L. Edelstein-Keshet,   
Science {\bf 284}, 99 (1999); M. Ballerini \textit{et al.},     
Proc. Natl. Acad. Sci. USA  {\bf 105}, 1232 (2008).
\bibitem{bird2} T. Feder,  
Phys. Today {\bf 60}, 28 (2007).



\bibitem{bact1} C. Dombrowski, L. Cisneros, S. Chatkaew,
 R.E. Goldstein, and J.O. Kessler,  
Phys. Rev. Lett. {\bf 93}, 098103 (2004); D. Volfson, S. Cookson, J. Hasty, and L.S. Tsimring, 
Proc. Natl. Acad. Sci. USA {\bf 105}, 15346 (2008).


\bibitem{cell1} P. Rorth, 
Trends Cell Biol. {\bf 17}, 575 (2007).



\bibitem{cytoskeleton2} T. Surrey, F. Nedelec, S. Leibler, and E. Karsenti,
Science {\bf 292}, 1167 (2001); P. Kraikivski, R. Lipowsky, and J. Kierfeld,  
Phys. Rev. Lett. {\bf 96}, 250813 (2006); V. Schaller, C. Weber, C. Semmrich, E. Frey, and A.R. Bausch,
Nature {\bf 467}, 73 (2010); X. Shi, Y. Ma,  
Proc. Natl. Acad. Sci. USA {\bf 107}, 11709 (2010).


\bibitem{vicsek1} T. Vicsek, A. Czirok, E. Ben-Jacob, I. Cohen, and O. Shochet,
Phys. Rev. Lett. {\bf 75}, 1226 (1995);
J. Toner and Y. Tu, Phys. Rev. Lett. {\bf 75}, 4326 (1995);
R.A. Simha and S. Ramaswamy, Phys. Rev. Lett. {\bf 89}, 058101 (2002);
G. Gregoire and H. Chate, Phys. Rev. Lett. {\bf 97}, 090602 (2006);
A. Baskaran and M.C. Marchetti, Phys. Rev. Lett. {\bf 101}, 268101 (2008);
A. Baskaran and M.C. Marchetti, Proc. Natl. Acad. Sci. USA {\bf 106}, 15567 (2008);
D. Saintillan and M. J. Shelley,  Phys. Rev. Lett. {\bf 100}, 178103 (2008);
J. Toner, Y. Tu, and S. Ramaswamy, Ann. Phys. (N.Y.) {\bf 318}, 170 (2005);
S. Ramaswamy, Annu. Rev. Condens. Matter Phys. {\bf 1}, 9.1 (2010).

\bibitem{actnem1} (a) S. Ramaswamy, R.A. Simda, and J. Toner, Europhys. Lett. {\bf 62}, 196 (2003); (b) S. Mishra and S. Ramaswamy, Phys. Rev. Lett. {\bf 97}, 090602 (2006);
(c) H. Chate, F. Ginelli, and R. Montagne, Phys. Rev. Lett. {\bf 96}, 180602 (2006);
(d) V. Narayan, N. Menon, and S. Ramaswamy, J. Stat. Mech. P01005 (2006);
 (e) V. Narayan, S. Ramaswamy, and N. Menon, Science {\bf 317}, 105 (2007);
(f) I.S. Aranson, D. Volfson, and L.S. Tsimring  LS, Phys. Rev. E {\bf 75}, 051301 (2007).

\bibitem{supp} See supplementary information for supporting materials and movies.


\bibitem{Shimada} T. Shimada, M. Doi, and K. Okano, J. Chem. Phys. {\bf 88}, 7181 (1988);
A. Ahmadi, M.C. Marchetti, and T.B. Liverpool, Phys. Rev. E {\bf 74}, 061913 (2006);
A. Baskaran and M.C. Marchetti, Phys. Rev. E {\bf 77}, 011920 (2008).


\bibitem{Lorenz} E.N. Lorenz, J. Atoms. Sci. {\bf 20}, 130 (1963).



\bibitem{Julicher} F. J\"{u}licher, A. Ajdari, and J. Prost, Rev. Mod. Phys. {\bf 69}, 1269 (1997).




\end{thebibliography}
\end{document}